\documentclass[twocolumn,amssymb,fleqn]{revtex4} 
\usepackage{epsfig,amssymb,amsmath,graphicx,subfigure,hyperref  }  
\usepackage{color}

\begin{document}

\title{Topological vacancies in spherical crystal clusters}
\author{Zhenwei Yao}
\affiliation{School of Physics and Astronomy, and Institute of Natural
Sciences, Shanghai Jiao Tong University, Shanghai 200240 China}
\begin{abstract} 
Understanding geometric frustration of ordered phases in two-dimensional
condensed matters on curved
surfaces is closely related to a host of scientific problems in condensed
matter physics and materials science. Here we show how two-dimensional
Lennard-Jones crystal clusters confined on a sphere resolve the geometric
frustration and lead to pentagonal vacancy structures.  These vacancies, 
originated from the combination of curvature and physical interaction, are found
to be topological defects and they can be further classified into dislocational
and disclinational types. We analyze the dual role of these crystallographic
defects as both vacancies and topological defects, illustrate their formation
mechanism, and present the phase diagram. The revealed dual role of the
topological vacancies may find applications in the fabrication of robust
nanopores. This work also suggests the promising potential of exploiting the
richness in both physical interactions and substrate geometries to create new
types of crystallographic defects, which have strong connections with the
design of crystalline materials.
     \end{abstract}
\maketitle

\section{Introduction}

Geometric frustration of ordered phases in two-dimensional condensed matters like crystals and liquid
crystals confined on curved surfaces can create
a myriad of emergent static~\cite{nelson2002defects, bausch2003grain,
vitelli2006crystallography, bowick2009two} and dynamic~\cite{keber2014topology,
marchetti2013hydrodynamics,irvine2012fractionalization, yao2016dressed} defect
structures. Through the Thomson problem of finding the ground state of
electrically charged particles confined on a sphere~\cite{thomson1904xxiv,
saff1997distributing,wales2006structure,bowick2007interstitial,bowick2007dynamics} 
and the generalized versions on typical curved surfaces~\cite{bowick2009two,koning2014crystals},
several defect motifs in crystalline order like isolated disclinations,
dislocations, scars and pleats have been identified in the past
decades.~\cite{irvine2010pleats, bowick2011crystalline,
kusumaatmaja2013defect,mehta2016kinetic,kelleher2017phase} As a fundamental
topological defect in crystalline order, a disclination refers to a vertex
whose coordination number $n$ is deviated from six in two-dimensional
triangular lattice, and a topological charge of $q=6-n$ can be assigned to an
$n$-fold disclination.~\cite{nelson2002defects} A dislocation is a topological
dipole composed of a pair of oppositely charged fivefold and sevenfold
disclinations. Scars and pleats are strings of fivefold and sevenfold
disclinations.~\cite{bowick2002crystalline, bausch2003grain, irvine2010pleats}
These frustration-induced defects are highly involved in many important physical
processes like 2D crystal melting,~\cite{kosterlitz1973ordering,
halperin1978theory,nelson1979dislocation,radzvilavivcius2012geometrical}
crystalline packings of twisted filament
bundles,~\cite{kamien1995iterated, kamien1996liquids, grason2012defects,azadi2012defects,bruss2013topological}
self-healing of crystalline order,~\cite{irvine2012fractionalization,
yao2014polydispersity} non-equilibrium behaviors in active
matters,~\cite{marchetti2013hydrodynamics, mognetti2013living,palacci2013living,
yao2016dressed} and materials design.~\cite{nelson2002towards,
devries2007divalent,brojan2015wrinkling, zhang2014defects,wales2014chemistry,
jimenez2016curvature} In a recent study, by creating depletion-induced
attraction between colloidal particles confined in the interior surfaces of spherical
water-in-oil droplets, vacancies are observed to proliferate in the resulting
spherical crystals.~\cite{meng2014elastic} While the disclination structure in
spherical crystals
has been extensively studied theoretically and experimentally,~\cite{nelson2002defects, chushak2005solid, bowick2009two,
wales2014chemistry,azadi2014emergent,azadi2016neutral} vacancies are generally
unstable in spherical crystals where particles interact by the long-range Coulomb potential.~\cite{bowick2007interstitial, bowick2007dynamics} The discovery of
the stable vacancies as a new defect motif in the curved crystals composed of attractive
colloidal particles raises a number of important questions, such as their
topological nature, the mechanism behind their formation, and their connection
with the fundamental topological defects of disclinations and dislocations.

\begin{figure*}[th]  
\centering 
\includegraphics[width=5.5in, bb = 150 1000 1500 1350]{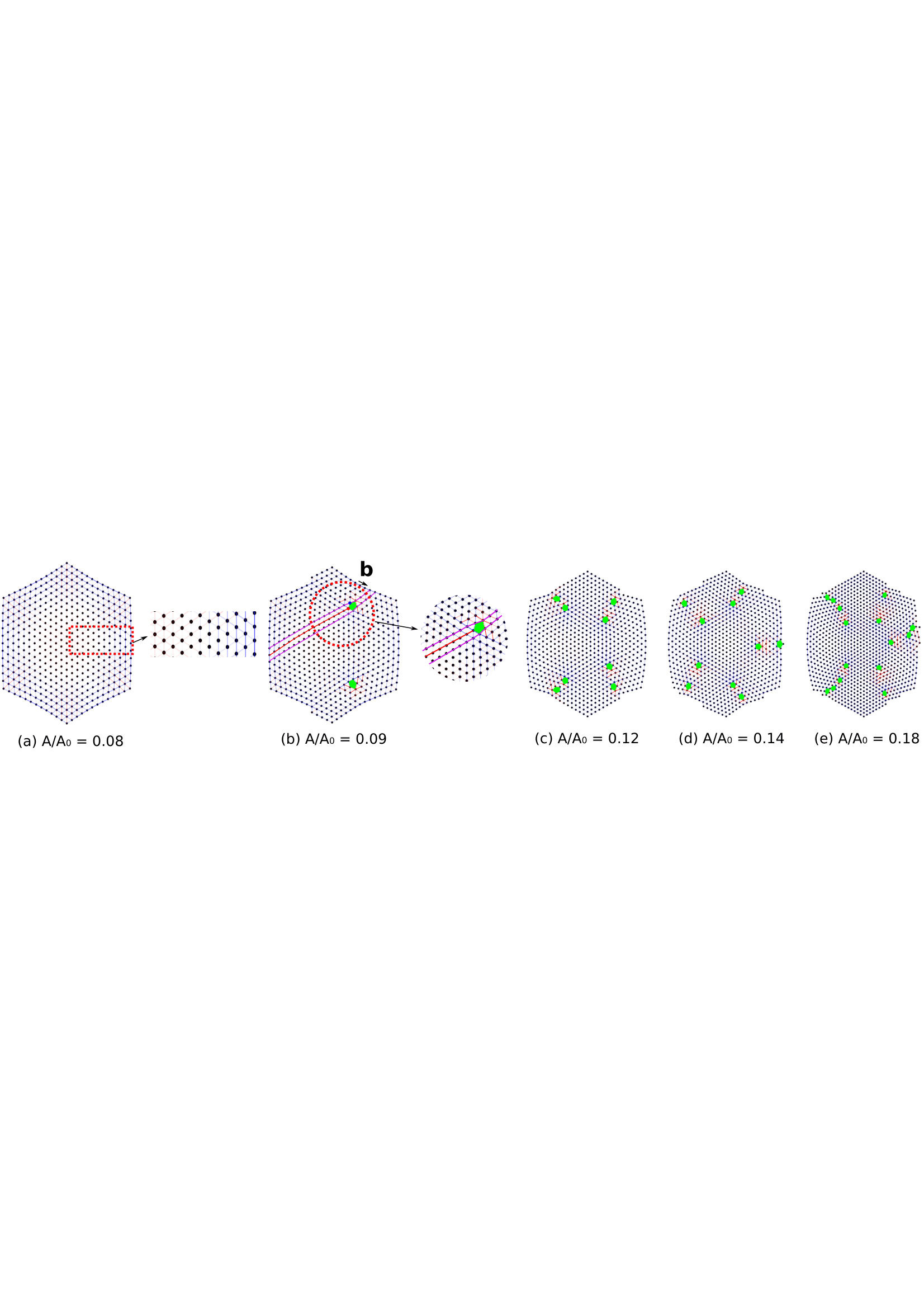}
\caption{Proliferation of pentagonal vacancies (in green) in crystal cluster on
  a sphere with accumulation of the curvature effect. The stress pattern
    indicated by colored bonds (red for stretching and blue for compression) is
    consistent with analytical elasticity theory (see text for more
	information). The plots are in the plane of the  spherical coordinates
    $\{\theta, \phi \}$. In each case, the coverage ratio of the crystal cluster
    $A/A_0$ is fixed. The inset in (b) is to illustrate the dislocational nature
    of the vacancy.  $R_0=20$.  }
\label{vac_I}
\end{figure*}

A model system suitable to address all these questions is to consider a
collection of point particles confined on a sphere interacting via the
Lennard-Jones (L-J) potential. The combination of depletion-induced short-range
attraction and hard-core repulsion
between colloidal particles seems responsible for the experimentally
observed vacancies in the rigid, two-dimensional colloidal crystals on spherical
droplets.~\cite{meng2014elastic} L-J potential possesses both attractive and
repulsive components and is a suitable form of interaction potential to
study the physics of vacancies.  The formally simple L-J potential has also been
extensively used to model various chemical and physical
bonds.~\cite{israelachvili2011intermolecular}  To lower the energy, the L-J
particles on a sphere are expected to spontaneously form geometrically frustrated
crystal clusters. We resort to numerical simulations in combination with
analytical theory to track the frustration of these spherical crystals and
analyze the resulting defect structures therein.  To highlight the geometric
effect and avoid mixing the curvature driven stress and the stress due to the
area constraint, we focus on particle clusters not fully occupying the sphere
and under free stress boundary condition.~\cite{roshal2013transfer} Note that
the model of L-J spherical crystal has been previously employed to study
underlying physics in several aspects of shell assembly like line tension, hole
implosion and closure catastrophe, where vacancy structure has been reported but
its topological property is still to be clarified.~\cite{luque2012physics}

In this work, we reveal the proliferation of {\it pentagonal} vacancies in
spherical crystal with the accumulation of curvature effect which are absent in
planar crystals. According to their topological nature, the pentagonal vacancies
can be classified into two categories: the type I dislocational and type II
disclinational vacancies. These two types of vacancies are topologically
equivalent to 
5-7 disclination pairs and 5-7-5 disclination strings, respectively.
Remarkably, both vacancies exhibit the identical pentagonal morphology despite
their distinct topological properties. Simulations show that their emergence is
controlled by the coverage ratio of the crystal cluster over the spherical
substrate. Geometric and elastic analysis shows the compatibility of the
pentagonal vacancies with spherical geometry. The revealed dual role of these
crystallographic defects as both vacancies and topological defects may be
exploited in the fabrication of nanopore structures in crystalline
membranes.~\cite{dinsmore2002colloidosomes,dubois2001self}  This work also
suggests the promising possibility of using the combination of physical
interaction and geometry to create rich types of defects that may find
applications in the design of crystalline materials.

\section{Model}

In our model the particles confined on the sphere of radius $R_0$ interact via the
L-J potential $V(r)=4\epsilon_0 \big[(\sigma_0/r)^{12} - (\sigma_0/r)^{6}
\big]$. The balance distance $r_m=2^{1/6}\sigma_0$. The initial configuration
of the spherical crystal cluster is prepared in the following way. We first
place a line of evenly distributed $M$ particles on the equator of the sphere,
and then pile particles layer by layer along the lines of latitude separated by
a distance of $\sqrt{3}r_m/2$ to fabricate a hexagonal crystal cluster. The
total number of particles in the constructed spherical cluster is
$N=\frac{1}{4}(3M^2+1)$. The initial configuration is free of stress in the
limit of $R_0 \rightarrow \infty$. We
employ the force method in simulations to track the evolution of the prepared
crystal cluster towards the lowest-energy state.~\cite{PhysRevE.88.012405, yao2013topological,
yao2016electrostatics} 
We first calculate the force on each particle by all
the other particles, and then simultaneously move all the particles by a small
distance of $s$ in each step until the energy of the system does not reduce any
more. Typically, $s=10^{-4}r_m$. The system evolves towards the lowest-energy state by repeating this
procedure.  The underlying topological defect structures in the spherical
crystals are analyzed by performing Delaunay triangulation with the software
Qhull.~\cite{barber1996quickhull} All the lengths are
measured in the unit of $r_m$. The controlling parameters are the radius of
the sphere $R_0$ and the coverage ratio $A/A_0$ of the crystal cluster. $A$ and
$A_0$ are the areas of the crystal cluster and the sphere, respectively. 
Note that in general a condensed matter system composed of many interacting
particles can be easily trapped in a metastable state.~\cite{chaikin00a,
  bruss2013topological} Furthermore, due
to the constraint imposed by the initial configuration of the fixed hexagonal
lattice, there is high probability that the resulting
lowest-energy particle configurations found in our simulations are not the true
ground states. However, it is observed that the defect
structures identified in the evolution of the crystal clusters can facilitate
the reduction of energy. It is therefore expected that the revealed topological
vacancies can characterize the basic defect state in the true
ground state.

\section{RESULTS AND DISCUSSION}

Figure~\ref{vac_I} shows the configurations of particles mapped to the
$\{\theta, \phi \}$ plane, where $\theta$ and $\phi$ are the polar and azimuthal
angles in spherical coordinates. $\theta \in [0,
\pi]$. $\phi \in [0, 2\pi )$. The stretching and compression of the bonds are
indicated by the red and blue lines, and more transparent
bonds assume less stress. A planar crystal can wrap a developable surface of zero Gaussian
curvature via pure bending without introducing any in-plane stress.~\cite{landau99a}  However,
the geometric incompatibility between a planar crystal and a spherical
substrate leads to the coupling of bending and in-plane
strain.~\cite{struik88a,audoly2010elasticity}
In other words, the originally planar crystal must modify its metric by
adjusting the bond length when it bends to fit the spherical geometry. The
distribution of the in-plane stress in the defect free configuration at
$A/A_0=0.08$ is shown in Fig.~\ref{vac_I}(a).  The basic
feature of the stress field is that the interior part of the crystal is
azimuthally stretched (the hoop stress $\sigma_{\theta\theta}>0$ for $r<r_c$),
the exterior part is azimuthally compressed ($\sigma_{\theta\theta}<0$ for
$r>r_c$), and the crystal is radially stretched everywhere (the radial stress $\sigma_{rr}
> 0$).

To understand the numerically observed stress pattern on the defect free
spherical crystal as shown in Fig.~\ref{vac_I}(a), we consider the elastic
deformation of the originally planar
crystal. The two-dimensional crystal cluster can be modelled as an
isotropic elastic sheet whose elastic properties are characterized by the
Young's modulus $Y$ and the bending rigidity $B$.  The ratio $t=\sqrt{B/Y}$ is
the effective thickness of the sheet. For simplicity, we consider a circular
elastic sheet of radius $r_0$. $r_0 << R_0$, and $r_0 >> t$, so
the stretching energy of the sheet dominates over the bending energy in the
deformation.~\cite{azadi2012defects,azadi2014emergent}
Therefore, it suffices to analyze the stress state over a planar elastic sheet. The effect
of the spherical substrate is to impose a normal force per area $P$ over the
elastic sheet and to pull
the sheet on its edge by the force $F_T$ per unit length to deform the sheet. The distribution of
the in-plane stress is governed by the force balance equations. With the boundary conditions of
$\sigma_{rr}(r=0) = \sigma_{\theta\theta}(r=0)$ and $\sigma_{rr}(r=r_0) = F_T$,
one can derive the analytical expressions for the in-plane stress
field:~\cite{grason2013universal} $\sigma_{\theta\theta}(r)/F_T=   \beta
\left( 1-3(r/r_0)^2 \right) /16  +  1$, and $\sigma_{rr}(r) /F_T = \beta \left(
1-(r/r_0)^2 \right)  /16 + 1$. The in-plane stress pattern is completely
controlled by the parameter $\beta=(r_0/R_0)^2(Y/F_T)$. The sheet is under pure
tension for $\beta<8$. For $\beta>8$, the hoop stress $\sigma_{\theta\theta}$
becomes compressive at the annulus $r\in (r_0\sqrt{(1+16\beta^{-1})/3}, r_0)$.
The coexistence of positive and negative $\sigma_{\theta\theta}$ regions in
the spherical crystal in Fig.~\ref{vac_I}(a) indicates that it falls in the regime
of $\beta>8$.

For larger spherical crystal clusters, we observe the proliferation of
vacancies as shown in Fig.~\ref{vac_I}(b)-\ref{vac_I}(e). A vacancy in the triangular lattice can be
identified where the local particle density
is significantly lower than the surrounding lattice. 
Remarkably, we notice
that these vacancies are pentagonal, which are 
different from the hexagonal vacancies in two-dimensional planar L-J
crystals.~\cite{yao2014dynamics} Geometric analysis reveals the dislocational
nature of these pentagonal vacancies.  The inset of Fig.~\ref{vac_I}(b)
illustrates the extra (red) line of particles terminated at the vacancy. The
integration of the resulting displacement field along {\it any} contour
enclosing the vacancy returns the constant Burgers vector $\vec{b}$ that
characterizes the dislocational nature of the pentagonal
vacancy.~\cite{landau99a} Delaunay triangulations show the dislocational nature
of the vacancies in Fig.~\ref{vac_I}(b); they can be represented by a pair of
fivefold [the red dot on the circled green pentagon in Fig.~\ref{vac_I}(b)] and
sevenfold [the blue dot] disclinations. Here, we note that in the spherical
crystals formed by long-range repulsive particles, vacancies and
interstitials are attracted and repelled by five-fold disclinations,
respectively.~\cite{bowick2007interstitial} Consequently, both vacancies and
interstitials typically break up when subjected to interactions with multiple
disclinations. These defects are therefore generally unstable, and have never
been found in equilibrium spherical crystals where particles interact by the
long-range Coulomb potential.~\cite{bowick2007dynamics}

\begin{figure}[t]  
\centering 
\includegraphics[width=3.2in]{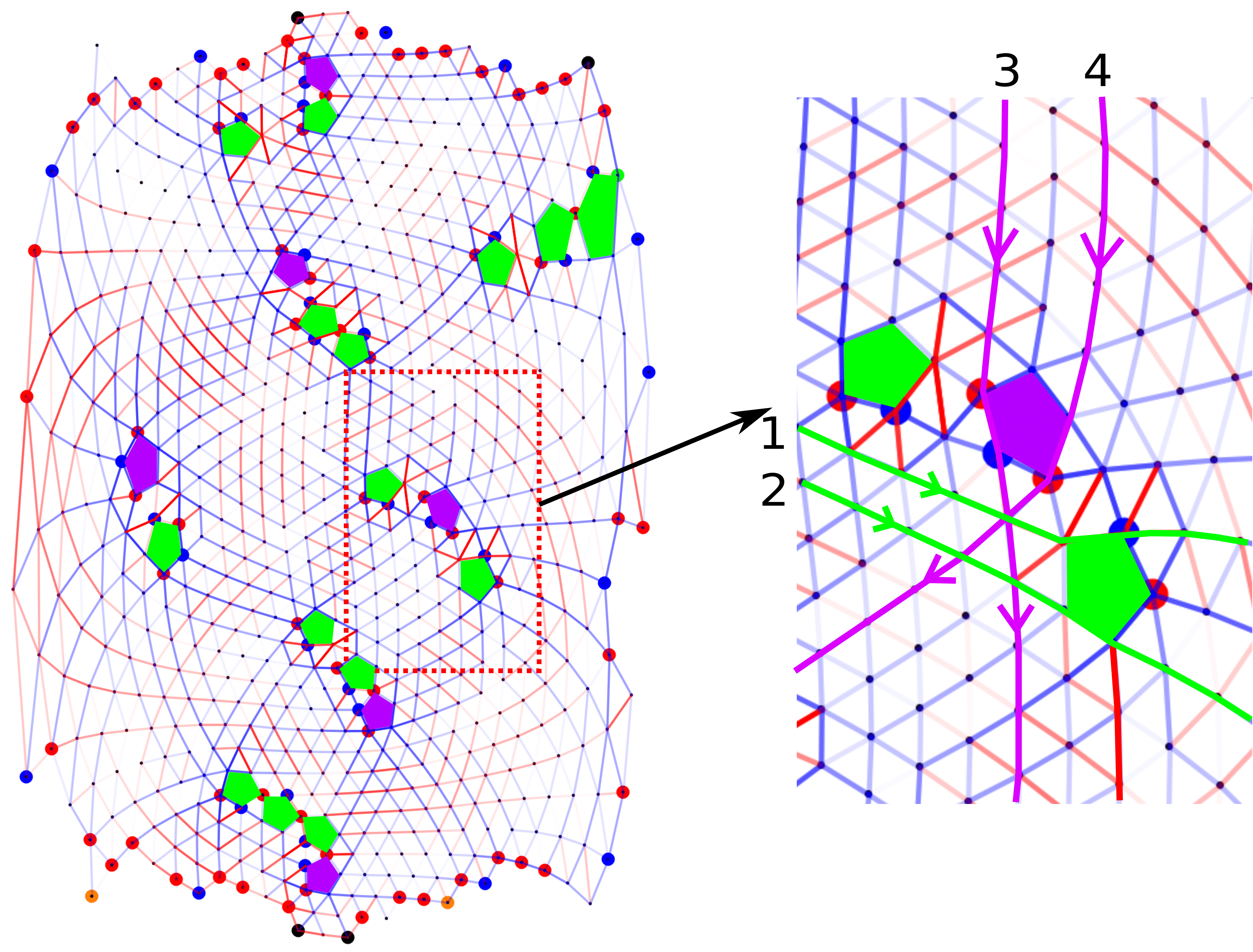}
\caption{Characterization of dislocational (green) and disclinational (purple)
  vacancies with the identical pentagonal morphology. The colored bonds indicate
    the stretching (red) and compression (blue) patterns; strain is smaller on
    more transparent bonds.  The plot is in the plane of the spherical coordinates
    $\{\theta, \phi \}$. $A/A_0=0.5$ $R_0=10$. 
   }
\label{vII}
\end{figure}

Comparison of the configurations in Fig.~\ref{vac_I}(a) and
Fig.~\ref{vac_I}(b) shows that the
pentagonal vacancies initially appear at the interface of the compression and
stretching zones. To account for these
preferred sites for the emergent defects, we examine the stress field of an isolated
dislocation in a two-dimensional isotropic elastic medium. In polar
coordinates, we have the stress field around an isolated dislocation:~\cite{landau99a}
\begin{eqnarray}
\sigma_{rr}=\sigma_{\theta\theta}=-\frac{bB\sin\theta}{r} \label{sigma}
\end{eqnarray}
and
\begin{eqnarray}
\sigma_{r\theta}=\frac{bB\cos\theta}{r},
\end{eqnarray}
where $B=\mu/2\pi(1-\nu)$, $\mu$ is the shear modulus, $\nu$ is the Poisson's
ratio, and $b$ is the magnitude of the Burgers vector. $\theta$ is measured
with respect to the direction of the Burgers vector. From eqn (\ref{sigma}), we
see that both radial and hoop bonds are stretched for $\theta \in (0,\pi)$
and compressed for $\theta \in (\pi,2\pi)$. Our simulations reveal stretched
red bonds and compressed blue bonds around the vacancy in the inset of
Fig.~\ref{vac_I}(b). Such a stress pattern is consistent with the above elastic
analysis based on the assumption of continuum elastic medium. The superposition
of this stress field on the stress pattern in Fig.~\ref{vac_I}(a) can
release both the compression in the exterior region [$\theta \in (0,\pi)$] and
the stretching in the interior region [$\theta \in (\pi,2\pi)$]. Therefore, the
pentagonal vacancies prefer to emerge at the interface of the compression and
stretching zones to release the elastic energy.

With the increase of the coverage ratio $A/A_0$,
Fig.~\ref{vac_I}(c)-\ref{vac_I}(e) show that the spherical crystal is plagued
with more dislocational pentagonal vacancies.  The arrangement of these defects
seems to follow some pattern. The distribution of the defects qualitatively
conforms to $p$-fold rotational symmetry. The value for $p$ increases from 2
(at $A/A_0=0.09$) to 5 (at $A/A_0=0.18$). Note that similar symmetric patterns
of boundary defect strings as well as their broken-symmetry states in the form
of forked-scar morphologies have been found in crystalline spherical caps under
external tension by the combination of
``free dislocation" simulations and
continuum elasticity theory. ~\cite{azadi2014emergent} The internal
dislocational defects in Fig.~\ref{vac_I}
also leave their signature in the exterior morphology of the crystal cluster;
we will show later that the steps on the contour of the crystal cluster
originate from the internal dislocations.  In Fig.~\ref{vac_I}(e), we notice
the unevenly distributed vacancies at the same defect string. The equally
placed dislocations constitute a grain boundary to separate
crystallites of distinct crystalline orientations.~\cite{chaikin00a} Here, the
spatial gradient in the density of dislocational vacancies will lead to net
topological charges according to the elasticity theory of topological
defects.~\cite{nelson2002defects}
Furthermore, the observed inhomogeneity in the arrangement of defects in
Fig~\ref{vac_I}(e) implies the appearance of disclinational defects at even
higher coverage ratios.

\begin{figure}[t]  
\centering 
\includegraphics[width=2.9in]{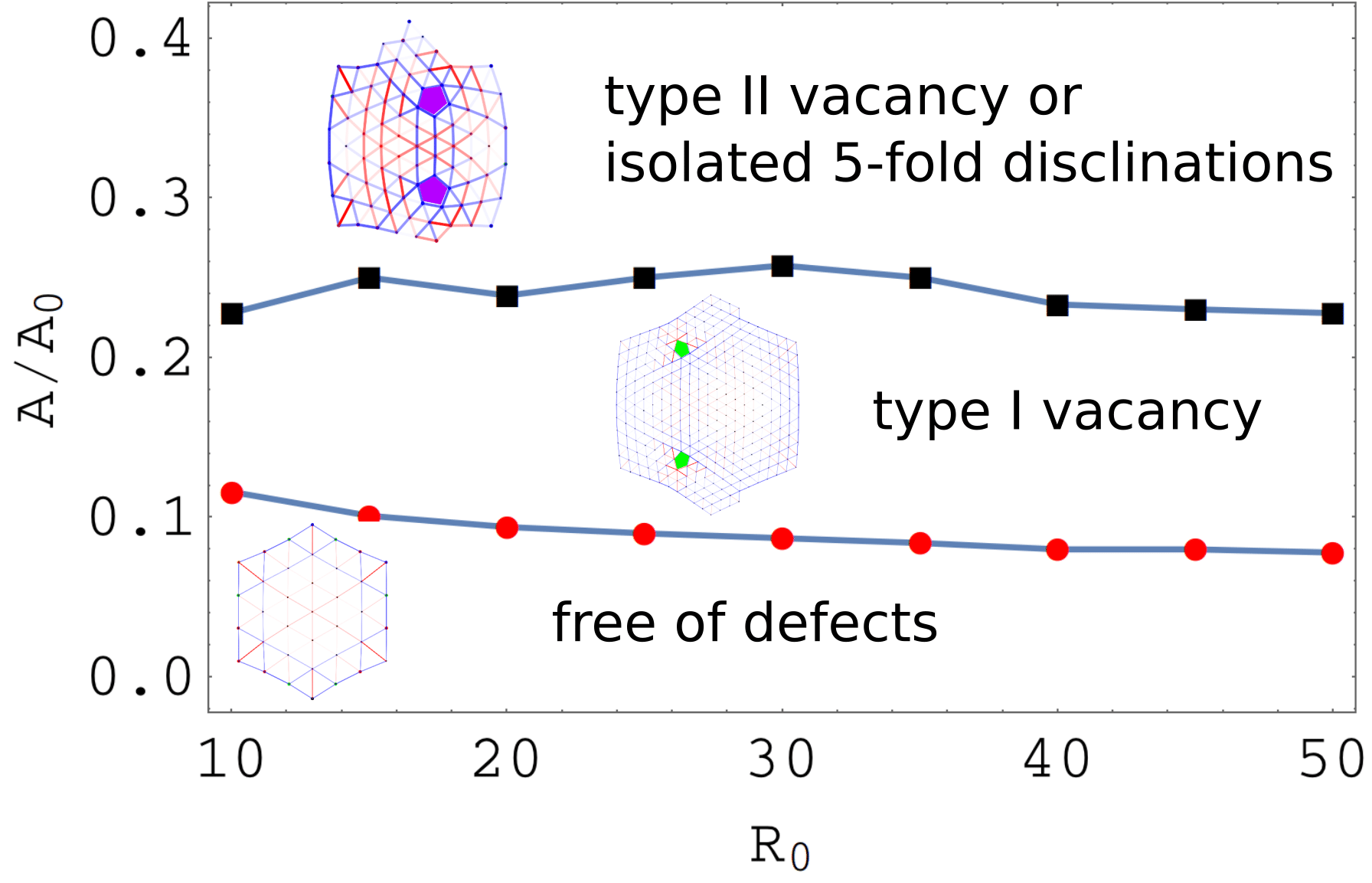}
\caption{Phase diagram of the defect state in L-J spherical crystals in the
  parameter space spanned by the coverage ratio $A/A_0$ and the radius $R_0$ of
    the sphere. The emergence of type I dislocational and type II disclinational
    vacancies is critically controlled by the coverage ratio of the
    crystal cluster over the sphere.   
}
\label{phase_diagram}
\end{figure}

Further increasing the coverage ratio leads to more pentagonal vacancies as
shown in Fig.~\ref{vII} for $A/A_0=0.50$. These vacancies can be classified by
their associated topological charge. Through Delaunay triangulation, we
identify a new type of pentagonal vacancy carrying positive topological
charge.  Such a vacancy is of disclinational nature, and is fundamentally
different from the previously discussed dislocational vacancy. In
Fig.~\ref{vII}, the two types of vacancies are indicated with different colors:
dislocational vacancies by green pentagons, and disclinational ones by purple
pentagons. The distinct nature of the pentagonal vacancies can also be
characterized by the deformation of the lattice lines. In the inset of
Fig~\ref{vII}, we illustrate that the lines 1 and 2 are still in parallel after
passing a type I dislocational vacancy; their separation is increased by a
Burgers vector's length.  In contrast, the originally parallel
lattice lines 3 and 4 are observed to bend and converge when passing a type II
disclinational vacancy, which resembles the convergence of light rays through a
convex mirror.~\cite{kamien2009extrinsic, irvine2010pleats}

We propose a geometric argument based on Gauss-Bonnet theorem to show that the
bending of the lattice lines around a type II vacancy is compatible with the
underlying spherical geometry.~\cite{struik88a} For a compact two-dimensional
Riemannian manifold $M$ with boundary $\partial M$, Gauss-Bonnet theorem states
that $\int _M K_G dA + \int_{\partial M} k_g ds = 2\pi \chi(M)$, where $K_G$ is
the Gaussian curvature, $k_g$ is the geodesic curvature, and $\chi(M)$ is the
Euler characteristic of the manifold $M$. Applying Gauss-Bonnet theorem on a
geodesic triangle $\triangle_{g.t.}$ formed by three geodesics on the sphere, we
obtain an elegant expression that connects the total curvature with the three
angles:~\cite{struik88a}
\begin{eqnarray}
\int_{\triangle_{g.t.}} K_G dA = \sum_{i=1}^{3} \theta_i -\pi, 
\label{KG_gt}
\end{eqnarray}
where the integration is over the geodesic triangle. It is straightforward to
check that the total
curvature of a planar triangle is zero.

Now let us consider a region enclosed by two converging geodesics over the sphere.
Bidirectional extension of two originally parallel geodesics on the sphere will
finally converge to form an eye-like shape. This region can be divided into two
geodesic triangles. From eqn (\ref{KG_gt}), we obtain the total curvature
\begin{eqnarray}
    \int K_G dA = \alpha_1 +   \alpha_2,  
    \label{int_KG}
\end{eqnarray}
where $\alpha_1$ and $\alpha_2$ are the angles at the two tips of the eye-like
region. The total curvature is smaller when the eye-like shape becomes sharper.
The limiting case is two parallel straight lines in a plane where the total
curvature vanishes. The convergence of the approximately geodesic lines $3$ and
$4$ in the inset of Fig.~\ref{vII} is therefore compatible with the positive
Gaussian curvature. Note that the convergence of lattice lines through a
type II vacancy is also seen in Volterra construction for
disclinations.~\cite{chaikin00a}  Specifically, a positive (or
negative) disclinational defect in two-dimensional triangular lattice can be
constructed by removing (or inserting) a wedge of angle $\pi/3$; this operation
leads to the convergence (or divergence) of originally parallel lines over a
planar disk.~\cite{irvine2010pleats}

\begin{figure}[t]  
\centering 
\includegraphics[width=3.2in]{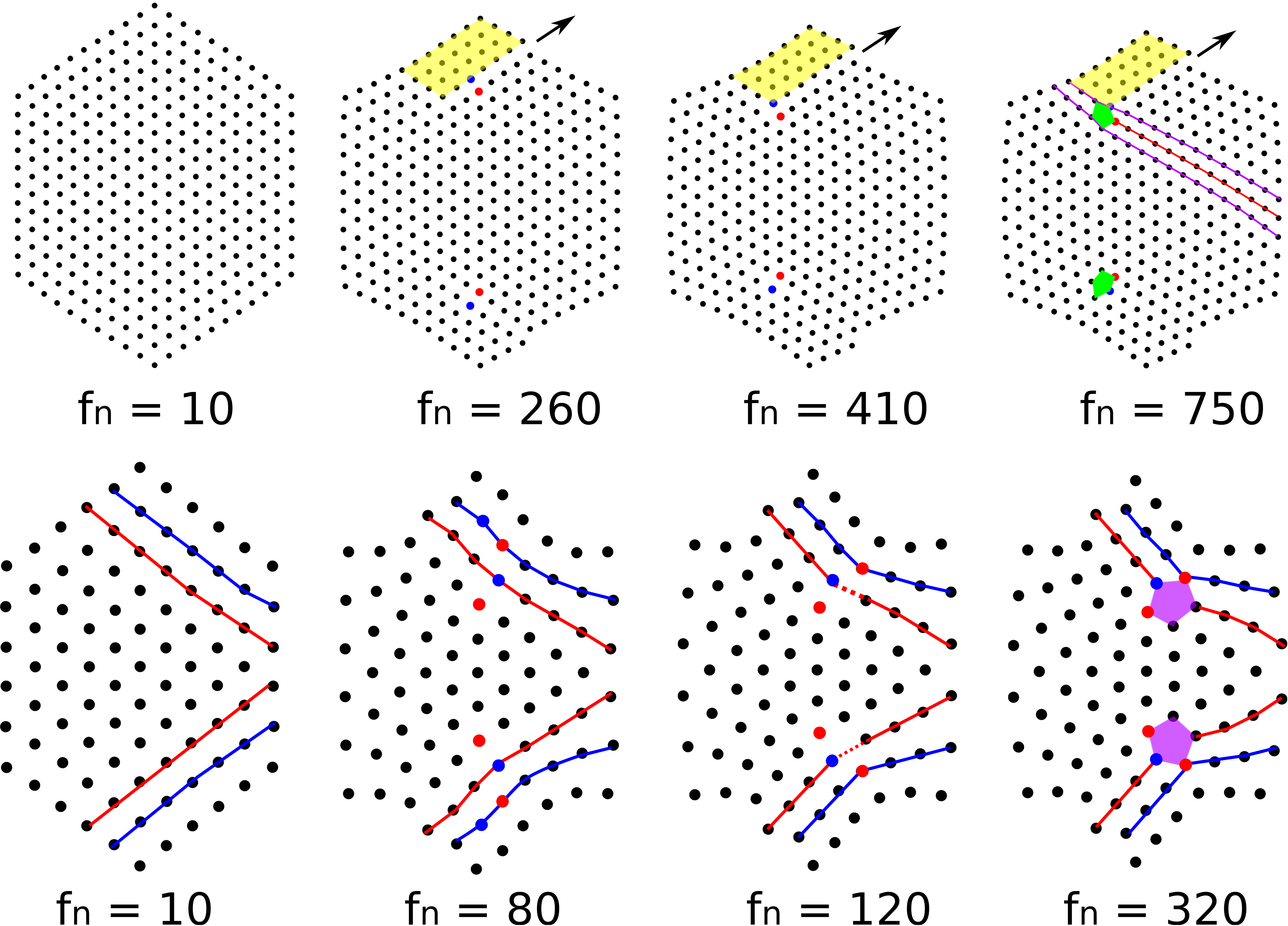}
\caption{Illustration of the distinct formation mechanisms of the dislocational
and disclinational vacancies by tracking the evolution of the spherical crystal
cluster towards the lowest-energy state. $f_n$ is the time frame. The green and
purple pentagons represent the resulting dislocational and disclinational
vacancies.  Upper figures: $A/A_0=0.1$.  $R_0=15$. Lower figures: $A/A_0=0.25$.
$R_0=5$.
}
\label{vac_origin}
\end{figure}

\begin{figure}[t]  
\centering 
\includegraphics[width=3.5in]{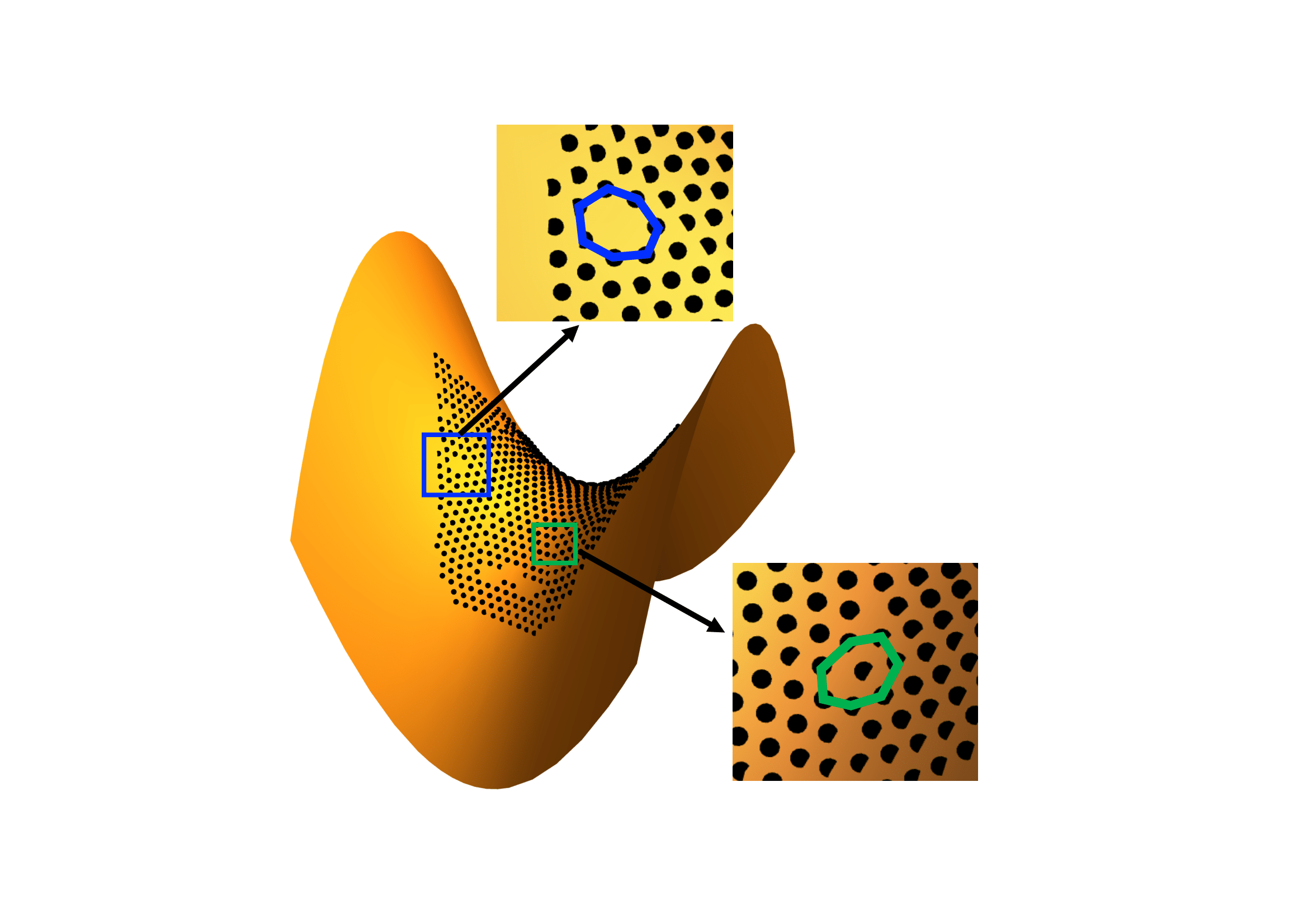}
\caption{Emergence of heptagonal vacancies (blue heptagon in the upper inset)
  and seven-fold disclinations (green heptagon in the lower inset) by confining
    a planar crystal cluster on the negatively curved hyperbolic paraboloid $z =
    (x^2-y^2)/16$. The negative-curvature driven coalescence of vacancies is
    also seen in the region below the lower green box. $N=721$.  }
\label{seven_vacancy}
\end{figure}

In Fig.~\ref{phase_diagram}, we present the phase diagram in the parameter
space of the coverage ratio $A/A_0$ and the radius $R_0$ of the sphere. It
clearly shows that the defect state of the spherical crystal is largely
controlled by the coverage ratio, and almost unaffected by the radius of
the sphere. Type I dislocational vacancies first emerge when $A/A_0$ exceeds about
$0.1$. With the increase of the coverage ratio, we observe the nucleation of
isolated dislocational vacancies into regularly arranged strings as shown in
Fig.~\ref{vac_I}. In addition, branched defect strings as well as cracks are
also numerically observed in some cases. Crack along these vacancy strings
occurs in the limit of vanishingly small separation between vacancies.
Fig.~\ref{phase_diagram} shows that further increase of the coverage ratio
above about $0.2$ leads to the proliferation of the type II pentagonal
vacancies or isolated fivefold disclinations within the crystal cluster. Here,
it is of interest to note that, in the crystalline packings of twisted filament
bundles, the emergence of dislocations and disclinations
in cross-sectional crystalline order of the bundles follows a similar phase
diagram in the regime of large bundles, where the appearance of these defects is
critically controlled by the accumulation of curvature
effect.~\cite{azadi2012defects}

We proceed to discuss the microscopic process in the formation of the
pentagonal vacancies.  In Fig.~\ref{vac_origin}, we present the typical
snapshots at different time frames labelled by $f_n$ in the evolution of the
crystal cluster with fixed coverage ratio towards the lowest-energy state. Here, a snapshot at "time
frame" $f_n$ refers to the particle configuration obtained after $n$ simulation steps in the
evolution of our system by the rule prescribed in the Model section.
In the upper snapshots in
Fig.~\ref{vac_origin}, we show that the formation of type I vacancies is
associated with the {\it collective} movement of particles in the yellow
shadow. This process can also be characterized in terms of the glide of the
dislocations.  The lower snapshots in Fig.~\ref{vac_origin} demonstrate the
formation of type II vacancies. At $f_n=80$, a pair of neutral linear defect
strings appear with the bending of the colored lattice lines. With the extension of
the red lattice line and further bending of the blue lattice line, the
sevenfold disclination originally sitting at the blue line vanishes at the
boundary and a pair of disclinational vacancies emerge at $f_n=320$.

We finally briefly discuss about which feature in the L-J potential is responsible for
the formation of the vacancies. Is it the local minimum structure or the steepness of the
physical potential? While the former feature is obviously responsible for the
condensation of particles to form a crystal, it is unknown if it also promotes
the emergent vacancies. To answer this question, we change the 12-6 potential to
the 4-2 potential $V_{4-2}(r)=4\epsilon_0 \big[(\sigma_0/r)^{4} -
(\sigma_0/r)^{2} \big]$. The $V_{4-2}(r)$ potential also possesses a local
minimum structure, but with a much softer ``ductility" than the 12-6 potential.
Specifically, the curvature of the 12-6 potential curve at its equilibrium
position is about 14 times than that of the 4-2 potential curve. By repeating the
simulations leading to the configurations in Fig.~\ref{vac_I}, we find that the
vacancy structures vanish. Instead, dislocations and disclinations appear to resolve the
geometric frustrations. Therefore, it is the steepness of the physical potential
that is responsible for the formation of the vacancy structures.

In the formation of the pentagonal vacancies in the spherical crystals, the featured
sufficiently steep energy minimum in the L-J potential curve can establish a bond
structure between neighboring particles, and
therefore support vacancies in both planar and curved
crystals.~\cite{meng2014elastic, yao2014dynamics} In simulations, we extend the
substrate geometry to hyperbolic paraboloid whose Gaussian curvature is
negative.  By confining L-J crystal clusters on such negatively curved surfaces,
we numerically observe heptagonal vacancies as well as seven-fold disclinations as
shown in Fig.~\ref{seven_vacancy}. We also notice the negative-curvature driven
coalescence of vacancies as shown in the region below the lower green
box in Fig.~\ref{seven_vacancy}. This phenomenon suggests that the model system
of negatively curved two-dimensional L-J crystal may be used to explore the
mechanism of crack formation. In contrast, under pure repulsive potentials,
vacancies created by removing particles have been observed to be filled by
neighboring particles to form dislocations.~\cite{pertsinidis2001diffusion} Note
that in our study we focus on behaviors of vacancies at zero temperature and
under the free stress boundary condition, where the particle density is largely
determined by the balance distance in the L-J potential. Robustness of vacancies
with the variation of both particle density (via imposing stress on the boundary
of a crystal cluster) and temperature has been shown  
in our previous study of planar L-J crystals, although the mobility of vacancies can be quantitatively
affected by particle density and temperature.~\cite{yao2014dynamics}  While it
is of interest to investigate the influence of temperature and particle density
on the behaviors of topological vacancies, this is beyond the scope of this
study.

\section{CONCLUSION}

In summary, we study the geometric frustration of Lennard-Jones spherical crystals, and
identify the pentagonal vacancies of dislocational or disclinational nature.  The
transition from the dislocational to the disclinational vacancies is largely
controlled by the coverage ratio of the crystal cluster. Clarification of the
topological nature of the pentagonal vacancies in this work enhances our
understanding about the basic defect structures in crystalline order on curved
surfaces. This work also suggests the promising potential of exploiting the
richness in both physical interactions and substrate geometries to create new
types of crystallographic defects. These defects may find applications in the
design of crystalline materials, notably in the fabrication of robust polygonal
nanopores on vesicles of spherical topology.

\section*{Acknowledgement}

This work was supported by NSFC Grant No. 16Z103010253, the SJTU startup fund
under Grant No. WF220441904, and the award of the Chinese Thousand Talents
Program for Distinguished Young Scholars under Grant No. 16Z127060004.

\end{document}